\documentclass[9pt,twocolumn,twoside]{article}

\usepackage{amsmath,amsthm,amssymb,amsfonts,mathtools,xcolor,authblk}
\usepackage[margin = 1cm]{geometry}

\def\boxit#1{\vbox{\hrule\hbox{\vrule\kern3pt
      \vbox{\kern3pt#1\kern3pt}\kern3pt\vrule}\hrule}}
\def\boxitd#1{\vspace{0.1in} \vbox{\hrule\hbox{\vrule\kern3pt
      \vbox{\abovedisplayskip=10pt \belowdisplayskip=10pt #1}
       \vrule}\hrule}}

\def\spb{\smallskip\par}

\def\bm#1{\mbox{\boldmath{$#1$}}}

\def\b0{\mbox{\bf 0}}

\newcommand{\norm}[1]{\left\lVert#1\right\rVert}

\def\Expect#1{\mathbb{E}\hspace{-0.03in}\left[ #1 \right]}

\def\Var#1{\mathbb{V}\hspace{-0.03in}\left[ #1 \right]}

\def\cor#1{\mathrm{cor}\hspace{-0.03in}\left(#1 \right)}
\def\cov#1{\mathrm{cov}\hspace{-0.03in}\left(#1 \right)}

\def\exp#1{\,\mathrm{exp}\hspace{-0.03in}\left(#1 \right)}

\def\Chi2#1{\,\Chi\hspace{-0.03in}\left( #1 \right)}

\newcommand\der[2]{\frac{d{#1}}{d{#2}}}

\newcommand\pder[2]{\frac{\partial #1}{\partial #2}}

\let\truesum=\sum
\def\sum{\mathop{\textstyle\truesum}\limits}

\def\_#1{{\underline{#1}}}
\def\'#1{{\mathbf{#1}}}
\let\truehat=\^
\let\^=\hat

\newcommand{\work}[2]{\ifnum\full=1{#1}\else{#2}\fi}

\def\vect#1{\bm{#1}}

\definecolor{dorange}{rgb}{1,.549,0}

\newcommand{\bang}[1]{\Big< #1 \Big>}

\newcommand{\la}[1]{\label{#1}}
\newcommand{\R}{\mathbb{R}}

\newcommand{\CP}{\mathbb{C}}
\newcommand{\p}{\partial}
\newcommand{\pp}[2]{\frac{\partial #1}{\partial #2}}
\newcommand{\ppn}[3]{\frac{\partial^{#1} #2}{\partial #3^{#1}}}

\newcommand{\LO}[1]{\boldsymbol\Delta #1}
\newcommand{\bfnabla}{\boldsymbol\nabla}

\newcommand{\defeq}{\vcentcolon=}

\newcommand{\tcb}[1]{\textcolor{blue}{#1}}
\makeatletter
\let\save@mathaccent\mathaccent
\newcommand*\if@single[3]{%
	\setbox0\hbox{${\mathaccent"0362{#1}}^H$}%
	\setbox2\hbox{${\mathaccent"0362{\kern0pt#1}}^H$}%
	\ifdim\ht0=\ht2 #3\else #2\fi
}
\newcommand*\rel@kern[1]{\kern#1\dimexpr\macc@kerna}
\newcommand*\widebar[1]{\@ifnextchar^{{\wide@bar{#1}{0}}}{\wide@bar{#1}{1}}}
\newcommand*\wide@bar[2]{\if@single{#1}{\wide@bar@{#1}{#2}{1}}{\wide@bar@{#1}{#2
}{2}}}
\newcommand*\wide@bar@[3]{
	\begingroup
	\def\mathaccent##1##2{
		\let\mathaccent\save@mathaccent
		\if#32 \let\macc@nucleus\first@char \fi
		\setbox\z@\hbox{$\macc@style{\macc@nucleus}_{}$}%
		\setbox\tw@\hbox{$\macc@style{\macc@nucleus}{}_{}$}%
		\dimen@\wd\tw@
		\advance\dimen@-\wd\z@
		\divide\dimen@ 3
		\@tempdima\wd\tw@
		\advance\@tempdima-\scriptspace
		\divide\@tempdima 10
		\advance\dimen@-\@tempdima
		\ifdim\dimen@>\z@ \dimen@0pt\fi
		\rel@kern{0.6}\kern-\dimen@
		\if#31
		
\overline{\rel@kern{-0.6}\kern\dimen@\macc@nucleus\rel@kern{0.4}\kern\dimen@}
		\advance\dimen@0.4\dimexpr\macc@kerna
		\let\final@kern#2
		\ifdim\dimen@<\z@ \let\final@kern1\fi
		\if\final@kern1 \kern-\dimen@\fi
		\else
		\overline{\rel@kern{-0.6}\kern\dimen@#1}
		\fi
	}
	\macc@depth\@ne
	\let\math@bgroup\@empty \let\math@egroup\macc@set@skewchar
	\mathsurround\z@ \frozen@everymath{\mathgroup\macc@group\relax}%
	\macc@set@skewchar\relax
	\let\mathaccentV\macc@nested@a
	\if#31
	\macc@nested@a\relax111{#1}
	\else
	\def\gobble@till@marker##1\endmarker{}%
	\futurelet\first@char\gobble@till@marker#1\endmarker
	\ifcat\noexpand\first@char A\else
	\def\first@char{}
	\fi
	\macc@nested@a\relax111{\first@char}
	\fi
	\endgroup
}
\makeatother

\title{Lagrangian Scaling Law for Atmospheric Propagation}

\author[1, *]{Sophia Potoczak Bragdon}
\author[2]{Daniel Cargill}
\author[3]{Jacob Grosek}

\affil[1]{Mathematics Department, Colorado State University, 1874 Campus Delivery, Fort Collins, CO 80523}
\affil[2]{Lockheed Martin Corporation, 199 Borton Landing Road, Moorestown, NJ 08057}
\affil[3]{Air Force Research Laboratory, Directed Energy Directorate, 3550 Aberdeen Ave SE, ABQ, NM 87117}

\affil[*]{Corresponding author's email: potoczak@math.colostate.edu}
\date{}

\begin{document}

\maketitle
\vspace*{-2cm}

\begin{abstract}
	A new scaling law model for propagation of optical beams through atmospheric 
	turbulence is presented and compared to a common scalar stochastic waveoptics technique.
	This methodology tracks the evolution of the important beam wavefront and phasefront 
	parameters of a propagating Gaussian-shaped laser field as it moves through atmospheric 
	turbulence, assuming a conservation of power.  
	As with other scaling laws, this Lagrangian scaling law makes multiple simplifying 
	assumptions about the optical beam in order to capture the essential features of interest, 
	while significantly reducing the computational cost of calculation.  
	This Lagrangian scaling law is shown to reliably work with low to medium turbulence 
	strengths, producing at least a $\sim2$x computational speed-up per individual 
	propagation of the beam and $>100$x memory reduction (depending on the chosen resolution).
\end{abstract}

\vspace*{-3mm}
\section{Introduction}
The modeling, analysis and simulation of optical wave propagation through the Earth's 
atmosphere is a challenging problem.  
This is mainly due to the presence of optical turbulence~\cite{kolmogorov1941local, 
frisch1995turbulence, andrews2005laser}, a term that refers to the stochastic multi-scale
variations in the index of refraction stemming from similar variations in density and temperature.  
As waves propagate, these refractive index perturbations induce corresponding phase perturbations, 
leading to scintillation, i.e. self-interference, and scattering, which broadens optical beams.
Compounding this problem further, is the stochastic nature of these perturbations, which introduces 
uncertainty that must be quantified to fully understand the problem of atmospheric 
propagation~\cite{goodman2015statistical, wang2015propagation, isaacs2018effect}.

In spite of the difficulties discussed above, there exists several approaches for simulating 
optical atmosphere propagation, each associated with a specific set of assumptions that tries 
to balance computational cost with model fidelity.  
On one end of this spectrum are waveoptics simulations, where the atmosphere is modeled as a 
random media using a prescribed probability distribution.  
Discretized realizations are drawn from this distribution and used as inputs to a stochastic 
partial differential equation (PDE) that models optical propagation.  
The PDE typically used is the reduced wave equation, also known as the paraxial Helmholtz equation.
This is derived from Maxwell's equations under the assumptions of small 
wavelength, i.e. in the optical regime~\cite{andrews2005laser}, and a high degree of coherency 
in the propagating optical wave.
The index of refraction parameter is modeled using a stochastic field and the resulting 
uncertainty is quantified through Monte-Carlo methods, where an ensemble of wave metrics 
are gathered and used to calculate statistics.  

At the other end of the modeling spectrum are scaling laws, a term referring to a set of formulas 
derived from analysis (asymptotic or numerical) of the aforementioned stochastic PDEs used in the 
waveoptics approach.  The objective in deriving these formulas is to map atmospheric statistics 
directly to statistics on the wave metrics without necessarily having to simulate the propagation 
directly, or to calculate the relevant statistics from the collected results of the ensemble of simulations.  
For example, the Rytov method gives a closed form representation for the first-order correction 
of a zeroth-order solution in the limit of small perturbations of the propagation 
medium~\cite{andrews2005laser, noriega2007rytov, wanjun2018propagation}.  
This first-order wave correction takes the form of an integral over both the zeroth-order solution 
and the medium perturbations along the propagation path.  
In cases where the zeroth-order solutions can be expressed in close form, e.g. Gaussian beams, 
these integrals can be well-approximated and used to form a mapping of atmospheric statistics to 
wave metrics~\cite{andrews2001theory, andrews2005laser, andrews2006strehl, noriega2007rytov}.  
It should be noted, however, that these derivations are always dependent on assumptions that 
limit the regime over which the resulting scaling laws are valid.  
The most common assumption is that the size of the perturbations are small, as in the Rytov 
example above, but could also include assumptions that neglect interactions between the wave 
and propagation medium as in case of thermal blooming~\cite{smith1977high, akers2019numerical}, 
a well known nonlinear interaction.  Nevertheless, these scaling law methods are commonly used 
to deliver first-order performance 
assessments of system design and deployment concepts~\cite{yura1971atmospheric, 
whiteley2010scaling, kitsios2012subgrid, van2013enhanced, shakir2016general}.

This paper introduces a new approach to forming a scaling law type approximation to the 
atmospheric propagation of optical beams based on a variational reformulation of the scalar 
stochastic paraxial Helmholtz equation commonly used in waveoptics simulations.  
Working with the Helmholtz equation in variational form allows approximations to be made 
through the use of suitably chosen trial functions, an approach that is commonly described 
as an extension of the Rayleigh-Ritz optimization procedure~\cite{fox1987introduction}.  
We begin by outlining a derivation of the paraxial Helmholtz equation and its reformulation 
as a variational problem.  
By introducing a trial solution in the form of parameterized Gaussian beam, one derives a 
set of stochastic ordinary differential equations (ODEs) that describe the evolution of these 
parameters as the beam propagates.  
A numerical results section shows how well this new scaling law performs 
in comparison to resolving the stochastic paraxial Helmholtz equation.  
To investigate the accuracy of the Lagrangian scaling law approximation to the paraxial Helmholtz 
equation (waveoptics approach), we compare the average of an ensemble of many realizations from 
both models to understand the effects of atmospheric turbulence. To do this, the two models are 
provided with equal initial conditions and are subjected to the comparable atmospheric turbulence conditions. 
The results show that the Lagrangian scaling law performs well for low to medium isotropic 
Kolmogorov turbulence conditions, greatly improving the computational cost of the calculation.

\vspace*{-3mm}
\section{Mathematical Formulation}

A propagating optical wave is described by Maxwell's equations.  
Since the atmosphere has virtually no magnetic susceptibility, one can 
capture the traveling wave by only tracking the electric field of the light.  
After a few manipulations of Maxwell's equations, one arrives at a modified wave 
equation for the electric field:  
\begin{equation}\la{equ:ModifiedWaveEquation}
	\LO{{\bf E}_{\ell}} - \bfnabla\left( \bfnabla \cdot {\bf E}_{\ell} \right) 
- \frac{1}{c^{2}}\ppn{2}{{\bf E}_{\ell}}{t} = \mu_{0} \ppn{2}{{\bf 
P}^{\ell}}{t},
\end{equation}
where the subscript $\ell$ specifies the angular frequency 
($\omega_{\ell}$) of the propagating wave, $c$ is the speed of causality, and 
$\mu_{0}$ represents the vacuum magnetic permeability.  
All interactions between the light and its medium are captured by the 
electric polarization term ${\bf P}^{\ell} = {\bf 
P}^{\ell}\hspace{-0.03in}\left( {\bf E}_{\ell} \right)$.  
The relevant interactions for atmospheric propagation through turbulence 
includes only the real-valued background mean index of refraction of the air 
($n_{\ell}^{0}$), a stochastic perturbation to this refractive index 
($\delta n_{\text{turb}}$), and a constant linear loss caused by absorption 
and/or scattering in the atmosphere ($\alpha_{\text{loss}}^{\ell}$).  
Loss is usually treated as a negative gain in the medium, and is derived as an 
imaginary perturbation to the refractive index~\cite{nagaraj20193d}.  
Mathematically, the electric polarization can be expressed as
\begin{equation*}
\begin{aligned}
{\bf P}^{\ell}\hspace{-0.03in}\left( {\bf E}_{\ell} \right) & \approx 
{\bf P}_{\text{background}}^{\ell}\hspace{-0.03in}\left( {\bf E}_{\ell} \right) 
+ {\bf P}_{\text{turb}}^{\ell}\hspace{-0.03in}\left( {\bf E}_{\ell} \right) + {\bf P}_{\text{loss}}^{\ell}\hspace{-0.03in}\left( {\bf E}_{\ell} \right) \\
{\bf P}_{\text{background}}^{\ell}\hspace{-0.03in}\left( {\bf 
E}_{\ell} \right) & \approx \varepsilon_{0} \Big( \left[ \big( {\bf 
n}_{\ell}^{0} \big)^{2} - {\bf 1} \right]{\bf E}_{\ell} \Big) \\
{\bf P}_{\text{turb}}^{\ell}\hspace{-0.03in}\left( {\bf E}_{\ell} 
\right) & \approx 2 \varepsilon_{0} \delta n_{\text{turb}} {\bf n}_{\ell}^{0}{\bf 
E}_{\ell} \\
{\bf P}_{\text{loss}}^{\ell}\hspace{-0.03in}\left( {\bf E}_{\ell} \right) & \approx \frac{i\alpha_{\text{loss}}^{\ell}\varepsilon_{0}c{\bf n}_{\ell}^{0}}{\omega_{\ell}}{\bf E}_{\ell}
\end{aligned} \quad \quad .
\end{equation*}
The vacuum electric permittivity is denoted as $\varepsilon_{0}$, and 
$\varepsilon_{0} \mu_{0} = c^{-2}$.   
In this model, the light propagates in +$z$-direction, ${\bf E}_{\ell} = {\bf 
E}_{\ell}({\bf r},t)$, ${\bf r} = (x,y,z)$, and ${\bf E}_{\ell} = 
\left[ E_{\ell}^{x} \ E_{\ell}^{y} \ E_{\ell}^{z} 
\right]^{\text{T}}$, where $[\cdot]^{\text{T}}$ is the transpose operator.  

It is assumed that the propagating light is highly coherent (from a laser 
source), and generally propagates in the +$z$-direction, making the 
$x,y$-directions transverse, which will be denoted with a $\perp$ symbol.  
Temporal coherence indicates that the light is near-monochromatic; other than 
the optical oscillation at the frequency $\omega_{\ell}$, the only other 
relevant timescales are that of the light travel time from the laser source 
to the target and that of the turbulence.  
This simulation assumes that the wavefront of the light can be propagated out 
to its endpoint (target) in a virtually static turbulence, since the turbulence 
changes over a much longer time period than the travel time.  
Spatial coherence indicates that the electric field can be well-approximated as 
a slowly varying envelope in the longitudinal direction that 
consistently oscillates in this direction at a frequency related to the 
wavenumber $k = n_{\ell}^{0}\omega_{\ell} / c$ of the field; this is known as 
the {\em paraxial approximation}.  
Moreover, omitting the small perturbation to the refractive index ($\delta n_{\text{turb}}$), 
the medium is mostly homogeneous, which means that Gauss's Law is applicable 
to this problem: $\bfnabla \cdot {\bf E}_{\ell} \approx 0$ since there is no volume charge density.
Additionally, it is assumed that the light is robustly linearly polarized in 
the $x$-direction, which means that $E_{\ell}^{y}$ and $E_{\ell}^{z}$ are 
negligibly small in comparison to $E_{\ell}^{x}$.  
Thus, 
\[
E_{\ell}^{x}({\bf r},t) \approx \text{Re}\left( A_{\ell}({\bf r}) \text{exp}\left[ ik^{\ell}z - i\omega_{\ell}t \right] \right),
\]
where Re$[\cdot]$ is the real-component-of operator, and $A_{\ell}({\bf r})$ 
is the slowly varying envelop of the electric field.  
With this ansatz, one derives from the wave equation~\eqref{equ:ModifiedWaveEquation} the {\em paraxial stochastic 
Helmholtz equation}:
\begin{equation}\la{equ:ParaxialStochasticHelmholtz}
\pp{A_{\ell}({\bf r})}{z} = \frac{i}{2k^{\ell}} 
\Delta_{\perp} A_{\ell}({\bf r}) + i k_{0}^{\ell} \delta n_{\text{turb}}({\bf 
r}) A_{\ell}({\bf r}) - \frac{\alpha_{\text{loss}}^{\ell}}{2} A_{\ell}({\bf r}),
\end{equation}
where $\Delta_{\perp} = \p^{2} / \p x^{2} + \p^{2} / \p y^{2}$ is the 
transverse Laplacian operator, and $k_{0}^{\ell} = \omega_{\ell} / c$.  
Based on the previous argument that the light propagates to its endpoint 
nearly instantaneously in comparison to any temporal changes to the turbulence, 
the paraxial stochastic Helmholtz equation is time independent, solved with a 
particular turbulence realization.  The statistics of the beam profile on target 
are found by solving this PDE multiple times, each with a new realization of the 
turbulence, which may be sampled from known statistics on $\delta n_{\text{turb}}({\bf r})$ 
that define the nature of the turbulence.  

This PDE can be nondimensionalized; first consider putting a ``hat'' $[\hat{\cdot}]$ 
over each parameter/variable in equation~\eqref{equ:ParaxialStochasticHelmholtz} in order to 
indicate that it has a dimension/unit.  
Next, make the following transformations: 
$\hat{x} = \widehat{l_{0}} x$, 
$\hat{y} = \widehat{l_{0}} y$, $\hat{z} = \widehat{L_{0}} z$, 
$\widehat{A_{\ell}} = \widehat{A_{0}} a$, 
$\widehat{\delta n_{\text{turb}}} = \sigma \delta n_{\text{turb}}$, 
$\xi = \widehat{l_{0}}^{2} \widehat{k_{0}^{\ell}} / \widehat{L_{0}}$, 
$\gamma = \widehat{l_{0}} \widehat{k_{0}^{\ell}} \sqrt{\sigma}$, 
and $\zeta = \widehat{l_{0}}^{2} \widehat{k_{0}^{\ell}} \widehat{\alpha_{\text{loss}}^{\ell}}$, 
where the unitless parameters/variables do not have ``hats'' over them.  
Moreover, for notational convenience, let $n \equiv n_{\ell}^{0}$.  
Note that $\sigma$ represents the strength of the atmospheric turbulence, 
and the stochastic perturbation to the refractive index ($\widehat{\delta n_{\text{turb}}}$) 
is actually unitless, but still uses a ``hat'' in order to distinguish it 
from its rescaling by $\sigma$.  
This yields
\begin{equation}\la{equ:NDHelmholtz}
  \begin{aligned}
    2i n \xi \pder{a}{z}({\bf r}) + \Delta_{\perp}a({\bf r}) + 
    2 n \gamma^{2} \delta n_{\text{turb}}({\bf r}) a({\bf r}) \ + \ & \\
    i n \zeta a({\bf r}) & = 0 \quad \quad .
  \end{aligned}
\end{equation}
Finally, $0 < \widehat{l_{0}} < \widehat{L_{0}}$ are length scales for the transverse and propagation dimensions
that may be directly related to the inner and outer scales of the atmospheric turbulence (e.g., Kolmogorov 
turbulence~\cite{kolmogorov1941local, frisch1995turbulence}), if one so chooses.  

Typically the wavelength of the laser light is chosen so that it transmits 
well through the atmosphere with low loss.  In fact, the loss parameter 
$\alpha_{\text{loss}}^{\ell}$ can be estimated using measured transmissions 
through the atmosphere~\cite{hemming2014review}.  Since the loss is a linear 
effect that occurs over the distance traveled\footnote{Note that the loss 
parameter $\alpha_{\text{loss}}^{\ell}$ has units of 1/m, indicating that its 
affect on the propagating wave grows with distance traveled.}, it has negligible 
transverse effects on the traveling wave, unlike the turbulence.  
Also, loss, or negative growth, directly attenuates the magnitude of the electric 
field amplitude, but does not alter the phase of wave, again unlike the turbulence.  
This means that it reasonable to treat the loss separately from the turbulence in 
the atmospheric propagation problem.  This separation can be accomplished by breaking 
the governing PDE~(\ref{equ:NDHelmholtz}) into two equations as follows:
\begin{align}
	\pp{a}{z}({\bf r}) & = -\frac{\alpha}{2} a({\bf r}) \la{equ:NDLossEq} \\
	\pp{a}{z}({\bf r}) & = \frac{i}{2 n \xi}\Delta_{\perp} a({\bf r}) + 
\frac{i \gamma^{2}}{\xi} \delta n_{\text{turb}}({\bf r}) a({\bf r}) \la{equ:NDHelmholtzNoLoss},
\end{align}
where $\alpha = \zeta / \xi = \widehat{L_{0}} \widehat{\alpha_{\text{loss}}^{\ell}}$.  

The loss equation~\eqref{equ:NDLossEq} can be solved analytically with an initial 
condition: $a(x,y,0) = a_{0} \phi(x,y)$, where $a_{0} \in \CP$ and $\phi(x,y)$ is the 
initial, real-valued (without loss of generality) transverse profile of the propagating 
wavefront, yielding $a({\bf r}) = a_{0} \phi(x,y) \exp{-\alpha z / 2}$ or 
$|a({\bf r})|^{2} = |a_{0}|^{2} \phi^{2}(x,y) \exp{-\alpha z}$.  
On the other hand, the stochastic PDE for turbulence~\eqref{equ:NDHelmholtzNoLoss} ought 
to conserve the energy/power in the wavefront as the light propagates; the beam 
may focus or spread out, change phase, or drift in the transverse direction from its 
original center position, but it will conserve energy/power as it propagates in the 
longitudinal direction.  Clearly, $|a_{0}|^{2}$, of the initial condition, is related to 
this energy/power within the beam, especially if the beam profile is normalized such 
that $\iint_{D_{\perp}} \phi^{2}(x,y) \ dxdy = 1$, where $D_{\perp}$ represents the 
transverse domain. The exponential factor $\exp{-\alpha z}$ attenuates that energy/power 
as the light propagates.  Therefore, since the solution to the lossless paraxial 
stochastic Helmholtz equation~\eqref{equ:NDHelmholtzNoLoss} reveals a conserved quantity, 
by attenuating that quantity according to the exponential factor, one captures the effect of atmospheric loss.  
The Lagrangian scaling law will focus on solving this paraxial stochastic Helmholtz equation~\eqref{equ:NDHelmholtzNoLoss} without loss.

\vspace*{-3mm}
\subsection{Variational Formulation}
The theoretical framework for the Lagrangian scaling law was inspired by D. 
Anderson’s works in~\cite{anderson1983variational, anderson1988approximate, anderson1996variational}, 
where Anderson used the ``variational approximation'' to simplify the resolution of a partial 
differential equation to the a lower-dimensional system. 
Lagrangian approach relies on {\em Hamilton's principle}, and ultimately needs 
the governing dynamical system to have a conserved quantity, and since 
atmospheric loss has already been separated out, the propagating optical wave 
ought to conserve energy/power.  

The Lagrangian approach to atmospheric propagation involves recasting the 
propagation equation in terms of critical points of a functional~\eqref{equ:NDHelmholtzNoLoss}.  
\begin{equation}\la{equ:Functional}
J\left( a, \nabla a \right) = \int_{0}^{L} \int_{-\infty}^{\infty} \int_{-\infty}^{\infty} {\cal L}_{\text{D}}\left( a, \nabla a \right) \ 
dxdydz,
\end{equation}
where ${\cal L}_{\text{D}}$ is the {\em Lagrangian density}.  
These critical points (points where derivative of the functional is zero) 
correspond to solutions of this PDE~(\ref{equ:NDHelmholtzNoLoss})
through corresponding {\em Euler-Lagrange equations}:
\begin{equation}
 \pder{\mathcal{L}_{D}}{a} - \sum_{i = 1}^{3}\pder{}{r_{i}}\left(\pder{\mathcal{L}_{D}}{(\partial_{r_{i}}(a))}\right) = 0,
\end{equation}
where $r_{i} \in \{ x, y, z \}$ for $i = 1,2,3$, respectively. 
For the lossless paraxial stochastic Helmholtz equation:
\begin{equation}\la{equ:LagrangianDensity}
 {\cal L}_{\text{D}}\left( a, \nabla a \right) = -2n \xi \, \mathrm{Im}\hspace*{-.1cm}\left(\bar{a} 
 \pp{a}{z}\right) - \Big| \nabla_{\perp} a \Big|^{2} +  2 n \gamma^{2} \delta n_{\text{turb}} \big| 
 a \big|^{2}.
\end{equation}
Here, $[\bar{\cdot}]$ denotes the conjugate operation.

\vspace*{-3mm}
\subsection{General Gaussian Ansatz}
The key to this method is the use of an ansatz, or trial solution, 
which defines the solution's dependence on a subset of the independent 
variables, and parameterizes the solution's dependence in the remaining 
independent variables. We illustrate the approach here with a Gaussian 
beam ansatz where we assume the solution is well represented in the 
transverse direction (with respect to propagation), 
i.e. the variables $x$ and $y$, by a Gaussian profile
\begin{equation}\la{equ:GaussianAnsatz}
 a\big( {\bf r}_{\perp}, {\bf p}(z) \big) = I\big( {\bf p}(z) \big) e^{-\Big( \Theta\big( {\bf r}_{\perp},{\bf p}(z) \big) + i \Phi\big( {\bf r}_{\perp}, {\bf p}(z) \big) \Big)},
\end{equation}
where
\begin{align*}
 I\big( {\bf p}(z) \big) & = \frac{C(z) \sqrt{W_{x}(z) W_{y}(z)}}{\sqrt{\pi}}, \\
 \Theta\big( {\bf r}_{\perp}, {\bf p}(z) \big) & = \frac{1}{2} \left( W_{x}^{2}(z) \big( x - X(z) \big)^{2} + W_{y}^{2}(z) \big( y - Y(z) \big)^{2} \right), \\
 \Phi\big( {\bf r}_{\perp}, {\bf p}(z) \big) & = P(z) + T_{x}(z) \big( x - X(z) \big) + T_{y}(z) \big( y - Y(z) \big) \ + \\
 & \hspace{13pt} F_{x}(z) \big( x - X(z) \big)^{2} + F_{y}(z) \big( y - Y(z) \big)^{2}, \text{ and} \\
 {\bf p}(z) & = \big[ C(z) \ W_{x}(z) \ W_{y}(z) \ T_{x}(z) \ T_{y}(z) \ X(z) \ Y(z) \\
 & \hspace{18pt} F_{x}(z) \ F_{y}(z) \ P(z) \big]^{\text{T}} \ \ .
\end{align*}
Note, this ansatz is parameterized through a set of real valued parameters 
(in ${\bf p}(z)$) which only depend on the independent variable $z$ 
representing length along the direction of propagation.  

The terms of the Gaussian ansatz can be mapped to beam characteristics.  
For example, $I\left( {\bf p}(z) \right)$ represents 
the peak of the beam amplitude, which depends on the parameters $W_{x}(z)$ and $W_{y}(z)$ --
representing the beam width in the $x$ and $y$ directions, respectively, 
and the parameter $C(z)$, associated with the total beam energy/power, i.e.
$\iint_{\R^{2}} |a\big( {\bf r}_{\perp}, {\bf p}(z) \big)|^{2} \ dxdy = C^{2}(z)$.  
Likewise, $\Theta\big( {\bf r}_{\perp} ,{\bf p}(z) \big)$ controls the beam profile 
through the width parameters, and through the parameters $X(z)$, $Y(z)$, which represent 
the (transverse) beam profile center position.  Finally, $\Phi\big( {\bf r}_{\perp} ,{\bf p}(z) \big)$ 
is the beam phase term, dependent on parameters for the piston $P(z)$, 
tip/tilt $T_{x}(z)$, $T_{y}(z)$ and focusing $F_{x}(z)$, $F_{y}(z)$.  Finally, 
note that the ansatz is completely determined in transverse direction, so all 
evolution of the solution is now determined through the parameters of this trial solution.  
Adding more parameters would give the ansatz more degrees of freedom in which to 
evolve and capture more of the dynamics of the true solution.  However, adding 
parameters arbitrarily could easily result in multiple parameters capturing the same 
evolution, while also severely complicating equations for this evolution discussed below.  
Rather, the parameters included in the ansatz should be, and, in this case, are, chosen carefully to reflect specific quantities and qualities of interest.
They are also chosen according to knowledge of how the dynamics of the some parameters affect other parameters of a realistic propagating optical beam.  
For example, when parameters that account for tilt in the phase, i.e. $T_{x}(z)$ and $T_{y}(z)$, are included in the ansatz, then the corresponding parameters that capture shifts in the beam center, i.e. $X(z)$ and $Y(z)$, should also be included in the trial solution, since the phase tilt parameters alter the position of the beam center.  
This is a matter of completeness of the model.  

Because the ansatz is completely determined in the transverse direction, the 
integrals over the transverse dimensions, i.e. the $x$ and $y$ variables, 
in relation~\eqref{equ:Functional} 
can be analytically evaluated, resulting in an averaged Lagrangian $\cal{F}_{\text{D}}$ for the 
dynamics captured by the parameterization
\begin{equation}\la{equ:Lagrangian}
   {\cal F}_{\text{D}}\left(\mathbf{p},\der{\mathbf{p}}{z} \right)
  = \int_{-\infty}^{\infty} \int_{-\infty}^{\infty} {\cal L}_{\text{D}}\left( \mathbf{r}_{\perp}, \mathbf{p}, \der{\mathbf{p}}{z} \right) \ d\hat{x} d\hat{y},
\end{equation}
while also redefining the functional in~\eqref{equ:Functional} 
in terms of just the evolution of the parameters in $z$,
\begin{equation}\la{equ:ReducedFunctional}
  J\left( \mathbf{p}, \der{\mathbf{p}}{z} \right) = \int_{0}^{L} {\cal F}_{\text{D}}\left( \mathbf{p},\der{\mathbf{p}}{z} \right) \ dz.
\end{equation}
Using the Gaussian ansatz in relations~\eqref{equ:LagrangianDensity}, \eqref{equ:GaussianAnsatz}, and 
\eqref{equ:Lagrangian} gives 
\begin{equation}\la{equ:ReducedLagrangianDensity}
 \begin{aligned}
  {\cal F}_{\text{D}}\left( {\bf p},\der{\bf p}{z} \right) & = 2n \xi C^2 \left(\der{P}{z} + \frac{\der{F_{x}}{z} }{2W_{x}^{2}} + \frac{\der{F_{y}}{z}}{2W_{y}^{2}} \right) \\
  &- C^{2} \left( \der{X}{z} T_{x} + \der{Y}{z} T_{y} \right) - \frac{C^2}{2} \left( W_{x}^{2} + W_{y}^{2} \right)\\
  &- C^{2} \left( T_{x}^{2} + T_{y}^{2} \right) - 2 C^{2} \left( \frac{F_{x}^{2}}{W_{x}^{2}} + \frac{F_{y}^{2}}{W_{y}^{2}} \right) \\
  &+  2n \gamma^2 \left< \delta n_{\text{turb}}, I^{2} e^{-2\Theta} \right>,
 \end{aligned}
\end{equation}
where 
\[
\bang{\cdot, *} \equiv \int_{-\infty}^{\infty} \int_{-\infty}^{\infty} \cdot * \ dxdy
\]
is an integration operation over the transverse domain, which can be 
interpreted as an innerproduct.  As long as the laser field maintains 
a Gaussian profile, the evolution of the beam should be well-described 
by the dynamics captured in relations~\eqref{equ:ReducedFunctional} and 
\eqref{equ:ReducedLagrangianDensity}.   
Furthermore, these dynamics are also described through the corresponding 
set of Euler-Lagrange equations given by 
\begin{equation}\la{equ:ReducedEulerLagrange}
  \pder{{\cal F}_{\text{D}}}{p_{j}} - \der{}{z}\pder{{\cal F}_{\text{D}}}{\left[ \der{p_{j}}{z} \right]} = 0
\end{equation}
for 
\[
p_{j} \in \Bigg\{ \underbrace{C}_{j = 1} \ \underbrace{W_{x}}_{j = 2} \ \underbrace{W_{y}}_{j = 3} \ \underbrace{T_{x}}_{j = 4} \ \underbrace{T_{y}}_{j = 5} \ \underbrace{X}_{j = 6} \ \underbrace{Y}_{j = 7} \ \underbrace{F_{x}}_{j = 8} \ \underbrace{F_{y}}_{j = 9} \ \underbrace{P}_{j = 10} \Bigg\}.
\]
Note, turbulence is introduced through stochastic term 
$2n \gamma^2 \left< \delta n_{\text{turb}}, I^{2} e^{-2\Theta} \right>$,
which can be interpreted as a projection of the stochastic 
index $\delta n_{\text{turb}}$ onto the term $I^{2} e^{-2\Theta}$, 
with $\gamma^2 = \left(\widehat{l_{0}} \widehat{k_{0}^{\ell}}\right)^2 \sigma$
being a nondimensionalized strength of the stochastic index 
variations.

\vspace*{-3mm}
\subsection{Governing Equations}
Applying the reduced Euler-Lagrange equation~\eqref{equ:ReducedEulerLagrange} 
to its corresponding reduced Lagrangian density~\eqref{equ:ReducedLagrangianDensity},
one derives
\newcounter{EquAlphCounter}
\setcounter{EquAlphCounter}{1}
\begin{align}
  C(z) & = C(0) \la{equ:LosslessAmplitudeMagnitude} \\ 
  \addtocounter{equation}{1}
  \der{W_{x}}{z}(z) & = \frac{2}{n \xi} F_{x}(z) W_{x}(z) \la{equ:InitialGEsLossless} \tag{\arabic{equation}\alph{EquAlphCounter}} \\ 
  \addtocounter{EquAlphCounter}{1}
  \der{W_{y}}{z}(z) & = \frac{2}{n \xi} F_{y}(z) W_{y}(z) \tag{\arabic{equation}\alph{EquAlphCounter}} \\ 
  \addtocounter{EquAlphCounter}{1}
  \der{T_{x}}{z}(z) & = -n \gamma^{2} \left< \delta n_{\text{turb}}({\bf r}), M_{T_{x}}({\bf r}) \right> \tag{\arabic{equation}\alph{EquAlphCounter}} \\ 
  \addtocounter{EquAlphCounter}{1}
  \der{T_{y}}{z}(z) & = -n \gamma^{2} \left< \delta n_{\text{turb}}({\bf r}), M_{T_{y}}({\bf r}) \right> \tag{\arabic{equation}\alph{EquAlphCounter}} \\ 
  \addtocounter{EquAlphCounter}{1}
  \der{X}{z}(z) & = -2T_{x}(z) \tag{\arabic{equation}\alph{EquAlphCounter}} \\ \addtocounter{EquAlphCounter}{1}
  \der{Y}{z}(z) & = -2T_{y}(z) \tag{\arabic{equation}\alph{EquAlphCounter}} \\ \addtocounter{EquAlphCounter}{1}
  \der{F_{x}}{z}(z) & = -\frac{W_{x}^{4}(z)}{2 n \xi} + \frac{2 F_{x}^{2}(z)}{n \xi} + \frac{\gamma^{2}}{\xi} \left< \delta n_{\text{turb}}({\bf r}), M_{F_{x}}({\bf r}) \right> \tag{\arabic{equation}\alph{EquAlphCounter}} \\ 
  \addtocounter{EquAlphCounter}{1}
  \der{F_{y}}{z}(z) & = -\frac{W_{y}^{4}(z)}{2 n \xi} + \frac{2 F_{y}^{2}(z)}{n \xi} + \frac{\gamma^{2}}{\xi} \left< \delta n_{\text{turb}}({\bf r}), M_{F_{y}}({\bf r}) \right> \tag{\arabic{equation}\alph{EquAlphCounter}} \\ 
  \addtocounter{EquAlphCounter}{1}
  \der{P}{z}(z) & = \frac{\left( W_{x}^{2}(z) + W_{y}^{2}(z) \right)}{2n \xi} - \frac{\gamma^{2}}{\xi} \left< \delta n_{\text{turb}}({\bf r}), M_{P}({\bf r}) \right>\tag{\arabic{equation}\alph{EquAlphCounter}}
\end{align}
where 
\begin{equation*}
  M_{T_{x}}({\bf r}) \defeq \frac{2}{C^{2}(z)} \pder{\big| a\big( {\bf r}_{\perp}, {\bf p}(z) \big) \big|^{2}}{X}
  = 4 W_{x}^{2} \left( x - X \right) \frac{|a|^{2}}{C^{2}} 
\end{equation*}
\begin{equation*}
  M_{T_{y}}({\bf r}) \defeq \frac{2}{C^{2}(z)} \pp{\big| a\big( {\bf r}_{\perp}, {\bf p}(z) \big)\big|^{2}}{Y} 
   = 4 W_{y}^{2} \left( y - Y \right) \frac{|a|^{2}}{C^{2}} 
\end{equation*}
\begin{equation*}
  M_{F_{x}}({\bf r}) \defeq \frac{W_{x}^{3}(z)}{C^{2}(z)} \pp{\big| a\big( {\bf r}_{\perp}, {\bf p}(z) \big)\big|^{2}}{W_{x}}
   = W_{x}^{2} \left[ 1 - 2 W_{x}^{2} \left( x - X \right)^{2} \right] \frac{|a|^{2}}{C^{2}}  
\end{equation*}
\begin{equation*}
  M_{F_{y}}({\bf r}) \defeq \frac{W_{y}^{3}(z)}{C^{2}(z)} \pp{\big| a\big( {\bf r}_{\perp}, {\bf p}(z) \big)\big|^{2}}{W_{y}}
   = W_{y}^{2} \left[ 1 - 2 W_{y}^{2} \left( y - Y \right)^{2} \right] \frac{|a|^{2}}{C^{2}} 
\end{equation*}
\begin{equation*}
\begin{aligned}
   M_{P}({\bf r}) &\defeq \frac{W_{x}(z)}{2 C^{2}(z)} \pder{\big| a\big( {\bf r}_{\perp}, {\bf p}(z) \big)\big|^{2}}{W_{x}} + \frac{W_{y}(z)}{2 C^{2}(z)} \pder{\big| a\big( {\bf r}_{\perp}, {\bf p}(z) \big)\big|^{2}}{W_{y}} + \\
  & \hspace{13pt} \frac{1}{2 C(z)} \pder{\big| a\big( {\bf r}_{\perp}, {\bf p}(z) \big)\big|^{2}}{C} \\
   & = \left[ 2 - W_{x}^{2} \left( x - X \right)^{2} - W_{y}^{2} \left( y - Y \right)^{2} \right] \frac{|a|^{2}}{C^{2}} 
\end{aligned}
\end{equation*}
Finally, the atmospheric loss due to absorption/scattering can be reintroduced 
by replacing the conservation of the amplitude magnitude relation~(\ref{equ:LosslessAmplitudeMagnitude}) with
\begin{equation}\la{equ:LossAmplitudeMagnitude}
C(z) = C(0)e^{-\frac{\alpha z}{2}},
\end{equation}
in accordance with the arguments concerning the loss relation~(\ref{equ:NDLossEq}).  

Therefore, the Lagrangian scaling law consists of nine coupled stochastic 
ODEs~(\ref{equ:InitialGEsLossless}$\tcb{\text{-}{\rm \alph{EquAlphCounter}}}$) 
that can either exclude atmospheric loss, using relation~(\ref{equ:LosslessAmplitudeMagnitude}), 
or include atmospheric loss, using relation~(\ref{equ:LossAmplitudeMagnitude}).  
Compare the fact that a numerical solver for the paraxial stochastic Helmholtz equation~\eqref{equ:NDHelmholtz} requires that one tracks a large number of 
discrete points in the transverse domain along the longitudinal propagation axis, 
whereas this new system requires that one tracks only nine parameters over the same distance.  
However, the Lagrangian scaling law still integrates over the transverse domain at every discrete longitudinal step.

\vspace*{-3mm}
\subsection{Gaussian Markov Approximations}
A typical assumption for the turbulence-induced perturbation to the refractive 
index is that it has a Gaussian probability distribution with a zero mean (or expected value) 
and a known standard deviation $\sigma_{\delta n}$:
\begin{equation*}
 \begin{aligned}
  \Expect{\widehat{\delta n}(\hat{\bf r})} & = 0, \quad & \Expect{\delta n({\bf r})} & = 0, \\
  \sqrt{\Var{\widehat{\delta n}(\hat{\bf r})}} & = \sigma_{\delta n}, \quad & \sqrt{\Var{\delta n({\bf r})}} & = 1,
 \end{aligned}
\end{equation*}
where $\widehat{\delta n} = \sigma \delta n$, 
and the subscript ``turb'' has been dropped for notational convenience.  
This choice offers another convenient property for the correlation between the perturbations 
of the refractive index at any two spatial positions:
\begin{align*}
  \cor{\delta n({\bf r}_{1}), \delta n({\bf r}_{2})} & \defeq \frac{\Expect{\prod_{i = 1}^{2} \left( \delta n({\bf r}_{i}) 
  - \Expect{\delta n({\bf r}_{i})} \right)}}{\prod_{i = 1}^{2} \sqrt{\Var{\delta n({\bf r}_{i})}}} \\
  & = \Expect{\delta n({\bf r}_{1}) \delta n({\bf r}_{2})}.
\end{align*}
Furthermore, it is common to assume that this perturbation to the index of refraction is 
delta-correlated in the propagation (longitudinal) direction of the light 
(referred to as the Markov Assumption~\cite{andrews2005laser}):
\[
\cor{\delta n({\bf r}_{1}), \delta n({\bf r}_{2})} = \cor{\delta n(x_{1}, y_{1}), \delta n(x_{2}, y_{2})} 
\delta_{\text{D}}(z_{1} - z_{2}),
\]
where $\delta_{\text{D}}$ represents the Dirac delta function.  

As indicated in the governing ODEs~(\ref{equ:InitialGEsLossless}$\tcb{\text{-}{\rm \alph{EquAlphCounter}}}$),
the evolution of the Gaussian beam contains continuous perturbations due to the overlap 
of the stochastic of refraction variations with, what will be called, 
Gaussian parameter modes ($M_j$): $\kappa_{\delta n, M} \vcentcolon= \left< \delta n, M \right>$.  
Since the modes ($M_j$) are deterministic, the mean of this overlap is
\[
\Expect{\kappa_{\delta n, M}} = \left< \Expect{\delta n}, M \right> = 0,
\]
In addition, the covariance between any two perturbations is given by
\begin{equation}\label{equation:Covariance}
 \begin{aligned}
  &\cov{\kappa_{\delta n, M_{1}}({\bf r}_{1}), \kappa_{\delta n, M_{2}}({\bf r}_{2})} = \\
  &\hspace{10pt}\delta_{\text{D}}(z_{1} - z_{2}) \iiiint_{-\infty}^{\infty} \cor{\delta n(x_{1}, y_{1}),
  \delta n(x_{2}, y_{2})} \ \cdot  \\
  &\hspace{15pt} M_{1}(x_{1}, y_{1}, z_{1}) M_{2}(x_{2}, y_{2}, z_{1}) \ dx_{1} dy_{1} dx_{2} dy_{2}
 \end{aligned}
\end{equation}
Note that $M_{1} \equiv M_{2}$ does not imply that ${\bf r}_{1} \equiv {\bf r}_{2}$, and vice-versa.  

Given these properties on the refractive index perturbation, the stochastic 
terms in~(\ref{equ:InitialGEsLossless}$\tcb{\text{-}{\rm \alph{EquAlphCounter}}}$) can be replaced by simple delta 
correlated Gaussian processes in the variable $z$ 
\begin{align}
 {\cal T}_{x}(z) = \left< \delta n_{\text{turb}}({\bf r}), M_{T_{x}}({\bf r}) \right> \\ 
 {\cal T}_{y}(z) = \left< \delta n_{\text{turb}}({\bf r}), M_{T_{y}}({\bf r}) \right> \\ 
 {\cal F}_{x}(z) = \left< \delta n_{\text{turb}}({\bf r}), M_{F_{x}}({\bf r}) \right>  \\ 
 {\cal F}_{y}(z) = \left< \delta n_{\text{turb}}({\bf r}), M_{F_{y}}({\bf r}) \right>  \\ 
 {\cal P}(z) = \left< \delta n_{\text{turb}}({\bf r}), M_{P}({\bf r}) \right>
\end{align}
with correlation matrix elements given by relation~\eqref{equation:Covariance}.  

This leads to a reduced version of the Lagrangian scaling law that is less computationally expensive when the stochastic terms in~(\ref{equ:InitialGEsLossless}$\tcb{\text{-}{\rm \alph{EquAlphCounter}}}$) are replaced by the five Gaussian processes defined above. This approximation of the stochastic terms eliminates the need to generate a set of perturbations to the index of refraction, $\delta n_\text{turb}$, and the subsequent integration over the transverse plane. In this simplified model, we instead draw realizations of each of the five Gaussian processes which is computationally cheap compared to the generation of $\delta n_\text{turb}$.

\vspace*{-3mm}
\section{Numerical Model Results}

To illustrate that the Lagrangian scaling law well-approximates the solution to the paraxial stochastic Helmholtz equation, the statistics, especially the average, of an ensemble of the beam propagation realizations from the solution to the Helmholtz equation is compared to the average result from the Lagrangian scaling law approach. Note that the ensemble statistics for the Lagrangian scaling law converge with fewer realizations compared to the waveoptics approach. However, in the results presented below, the ensemble statistics are computed for 400 realizations of both models.
For these comparisons, the Lagrangian scaling law and the paraxial Helmholtz equation are supplied with the same initial conditions, and the perturbation to the index of refraction are randomly sampled using the same statistical characteristics.  
The realizations of the index of refraction at any given discrete longitudinal point are also called {\em phase screens}, and they are generated using the {\em circulant embedding} method outlined in~\cite{circulantembedding}. 

The stochastic paraxial Helmholtz equation (the waveoptics approach) is solved via a Strang-splitting (split-step) scheme in which the stochastic term is treated separately from the diffusive term. 
A Strang-splitting scheme is a standard numerical method for solving partial differential equations, including the paraxial Helmholtz equation~\cite{strang1968split, mcnamarastrangchapter}. 
In this model, the transverse plane is equipped with periodic boundary conditions; however, the transverse domain is always chosen large enough that the beam does not substantially encounter these periodic boundaries.  
The diffusive term is treated with the fast Fourier transform (FFT) algorithm, and the stochastic term is viewed as a phase contribution for each particular realization of the phase screen ($\delta n_\text{turb}$). 

For simplicity, the governing set of stochastic ODEs corresponding to the Lagrangian scaling law is solved via the backwards Euler implicit method.  
It is important to note that the Lagrangian scaling law model is susceptible to convergence issues when using an explicit finite difference scheme.  
It is not yet clear whether the Lagrangian scaling law can be successfully implemented for strong turbulences because the numerical scheme seems to go unstable in such cases.  
This may be correctable with a more suitable numerical method -- further testing is needed.

\vspace*{-3mm}
\subsection{Atmospheric Turbulence Statistics}

Optical turbulence is typically represented through the Kolmogorov model~\cite{andrews2005laser} in which the stochastic variations of the index of refraction are described by the structure function:
\begin{equation*}
	\begin{split}
		& \Expect{(\delta n(\vect{r}_{1,\perp}) - \delta n(\vect{r}_{2,\perp}))^2} = 2 \left( 1 - \cor{\delta n(\vect{r}_1), n(\vect{r}_2)} \right) \\
		& = \frac{\widehat{C_n}^2 \widehat{l_{0}}^{\frac{2}{3}}}{\sigma_n^2}
			\begin{cases}
				\left|\vect{r}_{1,\perp} - \vect{r}_{2,\perp}\right|^{2}, & \text{for } 0 < \left| \vect{r}_{1,\perp} - \vect{r}_{2,\perp} \right| \leq 1 \\
				\left|\vect{r}_{1,\perp} - \vect{r}_{2,\perp} \right|^{\frac{2}{3}}, & \text{for } 1  < \left| \vect{r}_{1,\perp} - \vect{r}_{2,\perp} \right| \leq  \frac{1}{\epsilon}
			\end{cases}
	\end{split} \, ,
\end{equation*}
where $\epsilon = \widehat{l_0} / \widehat{L_0}$, and $\widehat{l_0}$ and $\widehat{L_0}$ are the characteristic length scales used to nondimensionalize the transverse and propagation directions, respectively.  
Thus, the correlation function is represented as
\begin{equation*}
	\begin{split}
		& \cor{\delta n_1,\delta n_2} = 1 \, - \\
		& \frac{\widehat{C_n}^2 \widehat{l_{0}}^{\frac{2}{3}}}{2 \sigma_n^2}  
		    \begin{cases}
				\left|\vect{r}_{1,\perp} - \vect{r}_{2,\perp} \right|^{2}, & \text{for } 0 < \left| \vect{r}_{1,\perp} - \vect{r}_{2,\perp} \right| \leq 1 \\
				\left|\vect{r}_{1,\perp} - \vect{r}_{2,\perp} \right|^{\frac{2}{3}}, & \text{for } 1  < \left| \vect{r}_{1,\perp} - \vect{r}_{2,\perp} \right| \leq \frac{1}{\epsilon}
			\end{cases} \, .
	\end{split}
\end{equation*}
If we assume that $|\cor{\delta n_1, \delta n_2}| \approx 0$ for $ \left| \vect{r}_{1,\perp} - \vect{r}_{2,\perp} \right| \approx 1/\epsilon$, then the variance of the index of refraction can be approximated as
\begin{equation*}
	\begin{aligned}
		\sigma_n^2 & \approx \frac{\widehat{C_n}^2 \widehat{L_0}^{2/3}}{2} \text{ and} \\
		\cor{n_1,n_2} & = 1 - 
		\begin{cases}
			\left|\vect{r}_1-\vect{r}_{2}\right|^{2}, & \text{for } 0 < \left|\vect{r}_1-\vect{r}_{2}\right| \leq 1 \\
			\left|\vect{r}_1-\vect{r}_{2}\right|^{2/3}, & \text{for } 1 \leq \left| \vect{r}_1 - \vect{r}_{2} \right| \leq  1/\epsilon
		\end{cases}
	\end{aligned} \, .
\end{equation*}
Note that $\sigma_n \equiv \sigma$ from the nondimensional parameters.  
These turbulence statistical properties are used in the numerical results presented hereafter for both the Lagrangian scaling law and the waveoptics model.

\vspace*{-3mm}
\subsection{Model Parameters}

For the numerical comparison the Lagrangian scaling law and the waveoptics model (scalar paraxial stochastic Helmholtz equation) are initialized as follows\dots \, First, the dimensional constants and characteristic scales are defined, and then the nondimensional counterparts are calculated.  The values of the physical constants used for the proceeding numerical results are given in Table~\ref{tab:physicalconstants}, and their corresponding scaled quantities are presented in Table~\ref{tab:scaledvars}.  
Since loss is not being considered: $\zeta = 0$. 
\begin{table}[b!]
	\centering
	\caption{The values of the physical constants and characteristic scales that describe the propagating laser field.}\label{tab:physicalconstants}
	\begin{tabular}{ccc}
		\hline
		physical quantity 		& symbol 								& value \\
		\hline \hline 
		wavelength 			& $\widehat{\lambda}$ 					& $10^{-6} \ \text{m}^{-1}$ \\
		inner-scale			& $\widehat{l_{0}}$ 					& $10^{-3} \ \text{m}$ \\
		outer-scale 			& $\widehat{L_{0}}$ 					& $10^{2} \ \text{m}$ \\
		index structure constant & $\widehat{C_n}$ 						& $10^{-9}  \ \text{m}^{-1/3}$ \\
		aperture diameter 		& $\widehat{D}$ 						& $2\cdot 10^{-2} \ \text{m}$ \\
		propagation distance 	& $\widehat{L_{z}}$ 					& $10^{4} \ \text{m}$ \\
		background index 		& $\hat{n}$ 							& $1 + 10^{-6}$ \\
		transverse length $x$ 	& $\widehat{L_{x}}$						& $0.5 \ \text{m}$ \\
		transverse length $y$ 	& $\widehat{L_{y}}$ 					& $0.5 \ \text{m}$ \\
		\hline
	\end{tabular}
\end{table}
\begin{table}[b!]
	\centering
	\caption{A listing of some scaled quantities and their corresponding values based on the given parameters of Table~\ref{tab:physicalconstants}.}\label{tab:scaledvars}
	\begin{tabular}{ccc}
		\hline
		computational quantity 	& symbol & value \\
		\hline
		\hline
		transverse length $x$ 	& $L_{x} = \widehat{L_{x}} /\widehat{l_{0}}$ 			& $500$ \\ 
		transverse length $y$ 	& $L_{y} = \widehat{L_{y}} / \widehat{l_{0}}$ 			& $500$ \\ 
		propagation distance 	& $L_{z} = \widehat{L_{z}} / \widehat{L_{0}}$ 			& $100$ \\
		scaled aperture 		& $D = \widehat{D} / \widehat{l_{0}}$ 					& $20$ \\
		index std. deviation & $\sigma = \frac{\widehat{C_n} \widehat{L_{0}}^{1/3}}{2}$ 	& $3.28 \cdot 10^{-9}$ \\
		wavenumber strength		& $\xi = \widehat{l_{0}}^{2} \widehat{k_{0}^{\ell}} / \widehat{L_{0}}$																				& $0.0623$ \\
		turbulence strength		& $\gamma = \widehat{l_{0}} \widehat{k_{0}^{\ell}} \sqrt{\sigma}$ 																					& $0.2599$ \\
		loss strength			& $\zeta = \widehat{l_{0}}^{2} \widehat{k_{0}^{\ell}} \widehat{\alpha_{\text{loss}}^{\ell}}$										& $0$ \\
		\hline
	\end{tabular}
\end{table}

\vspace*{-3mm}
\subsection{Initial Conditions}

\begin{figure}[t!]
	\centering
	\includegraphics[scale=0.5]{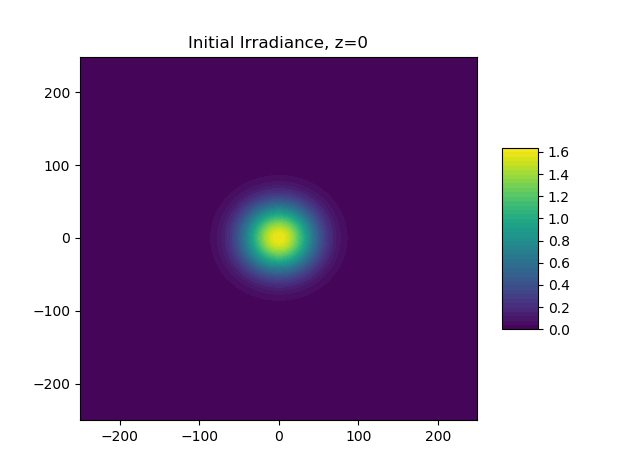}
	\caption{The initial irradiance for both the Lagrangian scaling law and the paraxial Helmholtz equation. }\label{fig:initialirradiance}
\end{figure}

The initial condition for both the Lagrangian scaling law and the waveoptics model will be given by the Gaussian ansatz~(\ref{equ:GaussianAnsatz}), using the ten parameters that describe the Gaussian.  
The particular choice of initial parameters is inspired by the deterministic Gaussian beam profile as described in the 2001 paper by Andrews et al.~\cite{andrews2001theory}.  
Specifically, these parameters are
\begin{equation}
	\begin{split}
		C(0) & = 100 = C \\
		W_x(0) & = W_y(0)=  \frac{w_0 \sqrt{n_0 \alpha}}{\sqrt{w_0^4 + (0 - z_w)^2}} \\
		T_x(0) & = T_y(0) = 0 \\
		X(0) & = Y(0) = 0 \\
		F_x(0) & = F_y(0) =  \frac{-0.5 n_0 \alpha(0 - z_w)}{w_0^4 + (0 - z_w)^2} \\
		P(0) & = -\arctan\Big(\frac{z_{w}}{w_0^2}\Big)
	\end{split} \, ,
\end{equation}
where $z_{w}$ is a specified scaled location and $w_{0}$ is the initial beam waist size. The location is prescribed to be $z_{w} = L_{z} / 2 = 50$, and the initial beam waist size is taken to be one quarter the diameter of the computational domain diameter: $w_{0} = D / 4 = 5$. 
Note that the beam width parameters in the Lagrangian scaling law are proportional to the inverse of the physical width of the beam. With this initial condition, any tilt in the system is strictly introduced through the interaction of the beam with the turbulent atmosphere. In the presence of no turbulence, i.e. $\delta n_\text{turb} = 0$, this initial condition choice allows us to know apriori the beam waist size and location. This is helpful for the case of weak atmospheric turbulence because we can expect that the beam waist size and location will be a perturbation away from the prescribed location in the initial condition. The initial irradiance is shown in Figure~\ref{fig:initialirradiance}. Note that the numerical results are presented using the nondimensional variables. 

\begin{figure}[t!]
	\centering
	\includegraphics[scale=0.5]{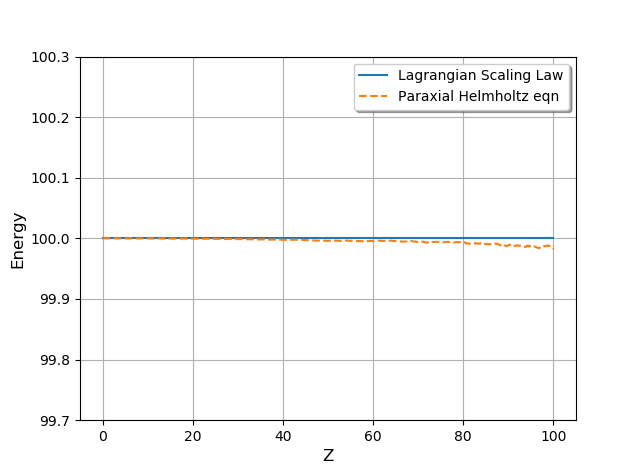}
	\caption{The beam power is conserved for both the Lagrangian scaling law and the paraxial Helmholtz equation, $C(z) = C(0)$ for all $z$. }\label{fig:energyconserved}
\end{figure}

From the derivation of the models, it is expected that the total beam power/energy is conserved throughout the propagation distance for both the Lagrangian scaling law and the paraxial Helmholtz equation. 
Thus, it is important to ensure the selected numerical methods for both models still conserve the total beam power. 
This can be easily checked by simply computing $C(z) \approx (\int_{-L}^{L}\int_{-L}^{L}\lvert a \rvert^2 dxdy)^{1/2}$ at each propagation step and we expect this to remain equal to the initial power of the beam. 
Figure~\ref{fig:energyconserved} shows the conservation of beam power over the propagation length for both models.

\vspace*{-3mm}
\subsection{Comparison}

To assess the accuracy of the Lagrangian scaling law in comparison to the waveoptics approach, an ensemble of 400 independent runs/realizations is conducted, and the average irradiance from \textcolor{black}{both} models are compared using the index structure constant value of $\widehat{C_n} = 10^{-9}$. 

\begin{figure}[t!]
	\centering
	\includegraphics[scale=0.5]{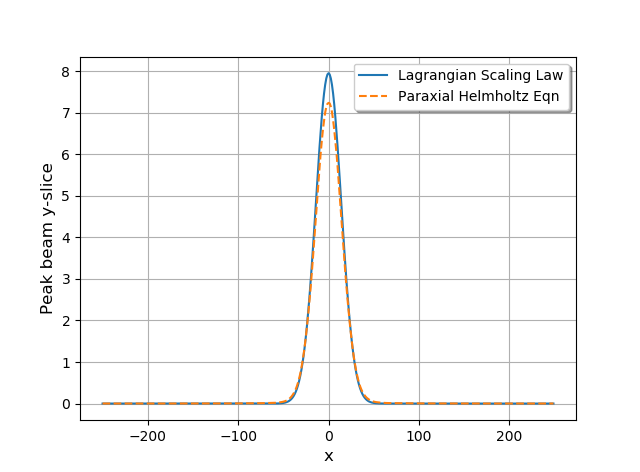}
	\caption{A comparison of a slice through the average irradiance along the line $x = 0$ at the location of the beam waist, $z = 50$, when the index structure constant is given by $\widehat{C_n} = 10^{-9}$.}\label{fig:irradiancexslice}
\end{figure}

To compare the difference in the value of the average peak irradiance from the two models we look at one-dimensional slices through the irradiance profile in both the $x$- and $y$-directions. Recall, the irradiance is found as the magnitude squared of the electric field. The errors are computed with the discrete $2$-norm as relative errors such that the waveoptics solution is considered to be trusted.  For notational convenience, let $I^{L}_{p}(x,y)$ be the peak irradiance from the Lagrangian scaling law solution and $I^{H}_{p}(x,y)$ be the peak irradiance from the paraxial Helmholtz equation. If we let $I(x,y)$ represent one of the above irradiances, then an $x$-slice through the irradiance is defined to be $I_{x}(y) = I(0,y)$ and a $y$-slice is defined to be $I_{y}(x) = I(x,0)$. 
A plot of the average peak irradiance $y$-slice is shown in Fig.~\ref{fig:irradiancexslice}. The relative error between the irradiance $x$-slices is $8.29 \cdot 10^{-2}$ and the relative error between the irradiance $y$-slices is $8.73 \cdot 10^{-2}$. 

\begin{figure}[t!]
	\centering
	\includegraphics[scale=0.5]{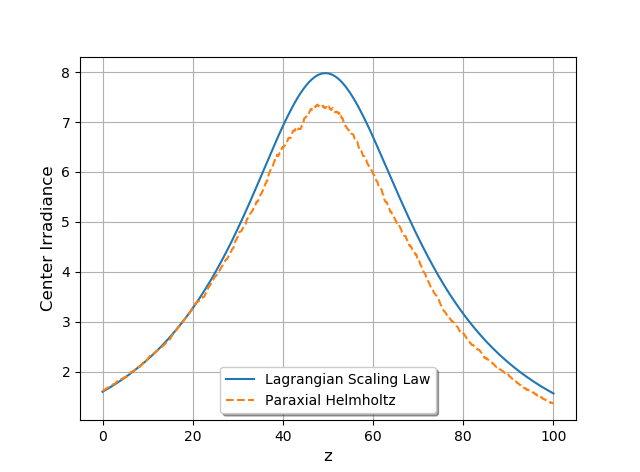}
	\caption{A comparison of the average center irradiance as a function of propagation distance when the index structure constant $\widehat{C_n} = 10^{-9}$.}\label{fig:centerirrcomp}
\end{figure}

Another measure for the accuracy is given by tracking the value of the peak irradiance. For the Lagrangian scaling law solution, this is simply given by the irradiance at the center of the Gaussian, which is at the mesh coordinates nearest to the $X = X(z)$ and $Y = Y(z)$ variables.  In the case of the paraxial Helmholtz equation, the location of the center irradiance is approximated numerically from the average of the ensemble.  The center irradiance of the Lagrangian solution will be denoted by $I^{L}_\text{center}(z)$, and the average center irradiance of the waveoptics solution will be denoted by $I^{H}_\text{center}(z)$.  The center irradiance is recorded for each propagation step, and again the relative error between the two models is measured in the $2$-norm:
\[
	\frac{\norm{I^{H}_\text{center}-I^{L}_\text{center}}_{2}}{\norm{I^{H}_\text{center}}_{2}} \, .
\]
The relative error for the peak irradiance along the propagation path is shown in Table~\ref{tab:avgcenterirr_errs}, and illustrated in Fig.~\ref{fig:centerirrcomp}. 
\begin{table}[b!]
	\centering
	\caption{The relative error of the average center irradiance between the paraxial Helmholtz equation and the 	Lagrangian scaling law.} \label{tab:avgcenterirr_errs}
	\begin{tabular}{ccc}
		\hline
		$\widehat{C_n}$ 		& error-type 		& $\frac{\norm{I^{H}_\text{center}-I^{L}_\text{center}}}{\norm{I^{H}_\text{center}}}$  \\ 
		\hline 
		\hline
		$10^{-9}$ 	& $2$-norm 		& $9.52 \cdot 10^{-2}$ \\
		$10^{-9}$		& $\infty$-norm 	& $1.10 \cdot 10^{-1}$ \\
		\hline
	\end{tabular}
\end{table}

\begin{figure}[t!]
	\centering
	\includegraphics[scale=0.5]{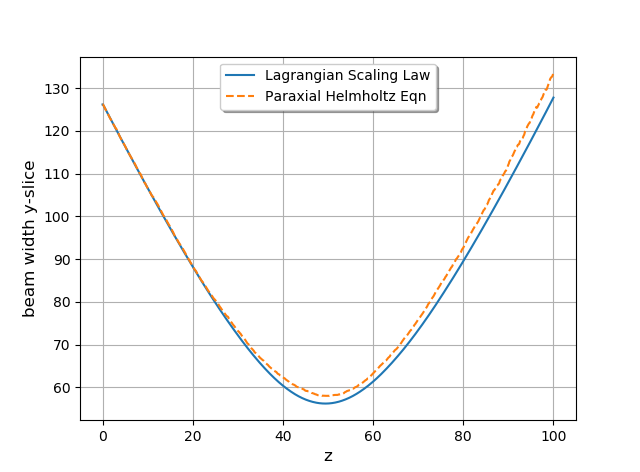}
	\caption{A comparison of the beam width as calculated from the centered $y$-slice through the irradiance profile along the entire propagation distance using the index structure constant $\widehat{C_n} = 10^{-9}$. The relative error between the beam width over propagation distance for the two models is $2.33 \cdot 10^{-2}$.}\label{fig:beamwidthcomp}
\end{figure}

Yet another metric of comparison is found in the width of the beam.
This width evolves as the light propagates, and the absolute peak irradiance over the entire propagation distance ought to correspond to the minimum beam width (the focal point). 
As a standard convention, the overall beam width for a given 1D slice through the Gaussian irradiance is bounded by the locations were the irradiance diminishes by a factor of $1/e^2$ from its peak.  
Again, only two slices centered on the $x$- and $y$-axes will be used for this calculation, producing a width value for each slice. 
Over the entire propagation distance, the relative error, measured in $2$-norm, of the beam width, calculated separately for the $x$- and $y$-slices through the irradiance profile, are $2.71 \cdot 10^{-2}$ and $2.33 \cdot 10^{-2}$, respectively.  
The evolution of this beam width for the $y$-slice is depicted in Fig.~\ref{fig:beamwidthcomp}.   
At the focal point of the waveoptics model, the relative error for the beam waist is given in Table~\ref{tab:avgbeamwaisterr}.  
\begin{table}[b!]
	\centering
	\caption{The relative error of the beam waist at the focal point of irradiance between the Lagrangian scaling law and the waveoptics model using $\widehat{C_n} = 10^{-9}$.}\label{tab:avgbeamwaisterr}
	\begin{tabular}{ccc}
		\hline 
		$\widehat{C_n}$ 	& peak irradiance slice 	& beam waist relative error \\
		\hline
		\hline 
		$10^{-9}$ & $x$-slice 			& $4.35 \cdot 10^{-2}$ \\
 		$10^{-9}$ & $y$-slice 			& $4.45 \cdot 10^{-2}$ \\ \hline
	\end{tabular}
\end{table}

%
%

As shown with this numerical example, the Lagrangian scaling law approximates the solution to the paraxial Helmholtz equation for this particular instance of weak turbulence.  A future effort of this work will include a survey of of numerical comparisons between the two models in the presence of stronger turbulence.

\vspace*{-3mm}
\section{Conclusions}

The Lagrangian scaling law offers a fast, reliable method for calculating the first-order approximation to the laser light atmospheric propagation problem.  
This may have many important directed energy beam control applications in the areas of scene generation, target 
detection, tracking, aimpoint maintenance, adaptive optics/atmospheric correction, et cetera. 
At this point, one run of the Lagrangian scaling law currently achieves $\sim2$x computation speed-up compared to one run of the waveoptics model. Further computational speed-up can be achieved by leveraging the simplifications outlined in the Gaussian Markov approximation subsection and this will be explored in future work. 
Most notably, the Lagrangian scaling law exhibits $>100$x memory reduction when compared to the waveoptics model. In the Lagrangian scaling law, the ten Gaussian parameters must be tracked over the course of propagation, on the other hand, the waveoptics model requires tracking the full transverse-field over the course of propagation.
The memory reduction achieved by the Lagrangian scaling law is dependent on the discretization parameters used in the waveoptics model. If one chooses a finer discretization of the transverse plane for the waveoptics model, then the memory reduction will increase.  
In the presence of weak atmospheric turbulence, the Lagrangian scaling law well-approximates the evolution of a Gaussian beam computed via the waveoptics model, as was shown with the above numerical comparisons. 
There are however some limitations to this Lagrangian scaling law approach.
For example, this is strictly a far-field approximation and does not account for (beam director) aperture obscuration.  

Though not explored in this effort, other trial solutions (non-Gaussian), e.g., the Zernike polynomial expansion, for beam profile could be explored in future efforts.
It is also worth noting that a vectorial Lagrangian approach ought to be feasible, where similar methodologies are applied to the full vectorial wave equation~\eqref{equ:ModifiedWaveEquation}.
Another avenue for future investigations would be to include thermal effects due to atmospheric heating 
by the laser beam in order to study the {\em thermal blooming} issue, especially within a control loop of a beam control system.  
Finally, it would be useful to complete a more formal comparison study of this Lagrangian scaling law to other existing scaling laws, effectively extending the work done by Bingham et al.~\cite{bingham2018wave}.

\vspace*{-5mm}
\section*{Acknowledgments}

The authors of this article would like to thank the U.S. Air Force Office of Scientific Research 
(AFOSR) Computational Mathematics program for providing funding for this project through project 
number 16RDCOR347, and to thank the Universities Space Research Association (USRA) for their funding 
and support through the Air Force Research Laboratory (AFRL) Scholars Program.  Additionally, the 
authors  wish to express gratitude to Olivier Pinaud of the Mathematics Department at Colorado State 
University, Dr. Laurence Keefe, NRC Senior Research Associate at the AFRL Directed Energy Directorate 
Laser Division (now with Zebara LLC), and Dr. Sami Shakir, Senior Scientist at Tau Technologies, 
for their impactful advise and suggestions, all of which substantially helped the progress this project.

{
\bibliographystyle{is-unsrt}
\bibliography{LSLForAtmosPropArticle.bib}
}

\end{document}